\documentclass[prl,aps,epsfig,twocolumn,showpacs]{revtex4}
\usepackage{epsfig,amssymb}

\newcommand\beq{\begin{equation}}
\newcommand\eeq{\end{equation}}
\newcommand\bea{\begin{eqnarray}}
\newcommand\eea{\end{eqnarray}}
\newcommand\non{\nonumber}
\newcommand\noi{\noindent}
\newcommand\al{\alpha}
\newcommand\be{\beta}
\newcommand\de{\delta}
\newcommand\ep{\epsilon}
\newcommand\La{\Lambda}
\newcommand\si{\sigma}
\newcommand\dg{\dagger}
\newcommand\up{\uparrow}
\newcommand\dn{\downarrow}
\newcommand\bib{\bibitem}

\begin{document}

\textheight=23.8cm

\title{\Large Renormalization group study of the Kondo problem 
at a junction of several Luttinger wires}
\author{\bf V. Ravi Chandra$^1$, Sumathi Rao$^2$ and Diptiman Sen$^1$} 
\affiliation{\it $^1$ Centre for High Energy Physics, Indian Institute of 
Science, Bangalore 560012, India \\
$^2$ Harish-Chandra Research Institute, Chhatnag Road, Jhusi, Allahabad 
211019, India}

\date{\today}

\begin{abstract}
We study a system consisting of a junction of $N$ quantum wires, where
the junction is characterized by a scalar $S$-matrix, and an impurity spin is 
coupled to the electrons close to the junction. The wires are modeled as weakly
interacting Tomonaga-Luttinger liquids. We derive the renormalization group 
equations for the Kondo couplings of the spin to the electronic modes on 
different wires, and analyze the renormalization group flows and fixed points 
for different values of the initial Kondo couplings and of the junction 
$S$-matrix (such as the decoupled $S$-matrix and the Griffiths $S$-matrix). We
generally find that the Kondo couplings flow towards large and antiferromagnetic
values in one of two possible ways. For the Griffiths $S$-matrix, we study one 
of the strong coupling flows by a perturbative expansion in the inverse of the 
Kondo coupling; we find that at large distances, the system approaches the
ferromagnetic fixed point of the decoupled $S$-matrix.  For the decoupled
$S$-matrix with antiferromagnetic Kondo couplings and weak inter-electron 
interactions, the flows are to one of two strong coupling fixed points in which
all the channels are strongly coupled to each other through the impurity spin.
But strong inter-electron interactions, with $K_\rho < N/(N+2)$, stabilize a
multi-channel fixed point in which the coupling between different channels
goes to zero. We have also studied the temperature dependence of the 
conductance at the decoupled and Griffiths $S$-matrices.

\end{abstract}
\maketitle
\vskip .6 true cm

\section{\bf I. Introduction}

The area of molecular electronics has grown tremendously in recent years as a 
result of the drive towards smaller and smaller electronic devices \cite{mol}. 
Molecular electronic circuits typically need multi-probe junctions. The first 
experimental growths of three-terminal nanotube junctions were not well
controlled \cite{zhou}; more recently, new growth methods have been developed 
whereby uniform $Y$-junctions have been produced \cite{li,kumar,terrones}. 
Transport measurements have also been carried out for the $Y$-junctions 
\cite{papad}, as well as for three-terminal junctions obtained by merging 
together single-walled nanotubes by molecular linkers \cite{chiu}.

On the theoretical side, there have been several studies of junctions of 
quantum wires. There have been detailed studies of carbon nanotubes with 
different proposed structures for the junction \cite{gan1,treboux}. Several 
groups have analyzed the geometry and stability of the junctions 
\cite{menon,meunier}. Junctions of quantum wires have also been studied 
\cite{nayak,chamon,lal1,das,chen,meden} in terms of one-dimensional wires, 
with the junction being modeled by a scattering matrix $S$. These studies 
include the effects of electron-electron interactions which are often cast 
in the language of Tomonaga-Luttinger liquid (TLL) theory 
\cite{gogolin,rao,giamarchi}.

Many earlier studies of junctions have only included `scalar' scatterings at 
the junction. i.e., the $S$-matrix has been taken to be spin-independent. The 
response of a junction of quantum wires to a magnetic impurity or an impurity
spin at the junction has recently been studied both experimentally 
\cite{leturcq} and theoretically \cite{rosch,oreg,pustilnik1,pustilnik2,ravi}.
As is well-known in {\it three} dimensions, an impurity spin can lead to the 
Kondo effect \cite{kondo}. [The coupling between the conduction electrons and 
the impurity spin grows as one goes to lower temperatures; this leads to a 
larger scattering and therefore a larger resistance as long as the temperature
is higher than the Kondo temperature $T_K$. Below $T_K$, the resistance due to
scattering from the impurity spin decreases (if the value of the impurity 
spin $\cal S$ is larger than or equal to half the number of channels $N$) 
because the spin decouples from the electrons.] The Kondo effect for a 
`two-wire junction' in a TLL wire has been studied by several groups 
\cite{lee,furusaki,fabrizio,frojdh,durga1,egger,kim,durga2}. Using a 
renormalization group (RG) analysis for weak potential scattering, Furusaki 
and Nagaosa showed that for an impurity spin of 1/2, there is a stable strong 
coupling fixed point (FP) consisting of two semi-infinite uncoupled TLL wires 
and a spin singlet \cite{furusaki}. For strong potential scattering, the above
FP is reached when the inter-electron interactions are weak. However, 
sufficiently strong inter-electron interactions are known to stabilize the 
two-channel Kondo FP instead \cite{fabrizio}. The Kondo effect has also been 
studied in crossed TLL wires \cite{karyn} and in multi-wire systems 
\cite{granath,yi}.

In this paper, we consider a junction of quantum wires which is characterized 
by an $S$-matrix at the junction; further, an impurity spin is coupled to the 
electrons at the junction. The wires are modeled as semi-infinite TLLs.
The details of the model defined in the continuum will be described in Sec. 
II. In Sec. III, we will discuss how RG equations for the Kondo couplings 
and for the $S$-matrix at the junction can be obtained by successively 
integrating out the electronic modes far from the Fermi energy. We find that 
the flow of the Kondo couplings involve the $S$-matrix elements, but the flow 
of the $S$-matrix elements do not involve the Kondo couplings (up to second 
order in the latter). To simplify our analysis, therefore, we concentrate on 
the FPs of the $S$-matrix RG equations and study how the Kondo couplings 
evolve in Sec. IV. For the case of $N$ decoupled wires, we find that for a 
large range of initial values of the Kondo couplings, the system flows to a 
multi-channel ferromagnetic (FM) FP lying at zero coupling. This FP is 
associated with spin-flip scatterings of the electrons from the impurity spin 
whose temperature dependence will be discussed. Outside this range, the flow 
is towards a strong antiferromagnetic (AFM) coupling. On the other hand, at
the Griffiths $S$-matrix (defined below), there is no stable FP for finite 
values of the Kondo couplings, and the system flows towards strong AFM 
coupling in two possible ways. We also consider the case when the scattering 
matrix has a chiral form. In this case, we find that the Kondo coupling matrix
for the three wire case has three independent degrees of freedom and a single 
FP at strong coupling.

The strong coupling flows will be further discussed in Sec. V where we will 
consider some lattice models at the microscopic length scale. As in the 
three-dimensional Kondo problem, we find that there are various possibilities 
depending on the number of wires $N$ and the spin $\cal S$ of the impurity, 
such as the under-screened, over-screened and exactly screened cases 
\cite{noz1}. 
We will generally see that a Kondo coupling which is small at high 
temperatures (small length scales) can become large at low temperatures (large
length scales). In Sec. VI, we will show that the vicinity of one of the strong
coupling FPs can be studied through an expansion in the inverse of the 
coupling; we will then find that the large coupling can be re-interpreted 
as a small coupling in a different model. 

In Sec. VII, we will study the case of decoupled wires with strong 
interactions using the technique of bosonization. Analogous to the results of 
\cite{fabrizio}, we find that the multi-channel ($N\ge 2$) AFM Kondo FP is 
stabilized for $K_\rho < N/(N+2)$. We will discuss the temperature 
dependence of the conductance in Sec. VIII at both high and low temperature;
we will compare the behaviors of Fermi liquids and TLLs. Sec. IX will contain 
some concluding remarks. A condensed version of some parts of this paper has 
appeared elsewhere \cite{ravi}.

Before proceeding further, we would like to emphasize that we have not used 
bosonization in this paper (except in Sec. VII), although this is a powerful 
and commonly used method for studying TLLs \cite{gogolin,rao,giamarchi}. In the
presence of a junction with a {\it general} scattering matrix, it is not known
whether the idea of bosonization can be implemented. (Some reasons for the 
difficulty in bosonizing are explained in Ref. \cite{lal1}). It is 
therefore necessary to work directly in the fermionic language. We have 
adopted the following point of view in this work \cite{lal1,yue}. We start 
with non-interacting electrons for which the scattering matrix approach and 
the Landauer formalism for studying electronic transport \cite{datta,imry} are
justified. We then assume that the interactions between the electrons are 
weak, and treat the interactions to first order in perturbation theory to 
derive the RG equations. This is the approach used in most of this paper.
Only in Sec. VII do we use bosonization to discuss the effect of strong
interactions for the case of decoupled wires, since that is one of the cases 
where bosonization can be used.

\section{\bf II. Model for several wires coupled to an impurity spin}

We begin with $N$ semi-infinite quantum wires which meet at one site which is 
the junction; on each wire, the spatial coordinate $x$ will be taken to 
increase from zero at the junction to $\infty$ as we move far away from the 
junction.

The incoming and outgoing fields are related by an $S$-matrix at the junction,
which is an $N\times N$ unitary matrix whose explicit values depend on the 
details of the junction. Hence the wave function corresponding to an electron 
with spin $\al$ ($\al = \up ,\dn$) and wave number $k$ which is incoming
in wire $i$ ($i=1,2,\cdots ,N$) is given by 
\bea
\psi_{i \al k} (x) &=& e^{-i(k+k_F)x} ~+~ S_{ii} e^{i(k+k_F)x} ~~{\rm 
on ~wire~} i ~, \non \\
&=& S_{ji} ~e^{i(k+k_F)x} ~~{\rm on ~wire~} j \ne i ~.
\label{wavefn}
\eea
Here $k$ is the wave number with respect to the Fermi wave number $k_F$ (i.e.,
$k=0$ implies that the energy of the electron is equal to the Fermi energy 
$E_F$). We will take $k$ to go from $- \La$ to $\La$, where $\La$
is a cut-off of the order of $k_F$; we will eventually only be interested in 
the long wavelength modes with $|k| \ll \La$. We will use a linearized 
approximation for the dispersion relation so that the energy of an electron 
with wave number $k$ is given by $v_F k$ with respect to the Fermi energy; 
here $v_F$ is the Fermi velocity, and we are setting $\hbar = 1$. In Eq. 
(\ref{wavefn}), we will refer to the waves going as $e^{-ikx}$ as the 
incoming part $\psi_{I i \al k}$, and the waves going as $e^{ikx}$ as the 
outgoing part $\psi_{O i \al k}$ or $\psi_{O j \al k}$. 

The second quantized annihilation operator corresponding to the wave function 
in Eq. (\ref{wavefn}) is given by
\beq
\Psi_{i \al k} (x) ~=~ c_{i \al k} ~\psi_{i \al k} (x) ~,
\eeq
where the wire index $i$ runs from 1 to $N$, and the total second quantized 
operator is given by
\beq
\Psi_\al (x) ~=~ \sum_i ~\int_{- \La}^\La ~\frac{dk}{2\pi} ~
c_{i \al k} ~\psi_{i \al k} (x) ~.
\eeq
(Note that it is not possible to quantize the system in terms of $N$ 
independent fields on each of the wires, because an electron that is incoming 
on one wire has outgoing components on all the other wires as well). The 
non-interacting part of the Hamiltonian for the electrons is then given by
\beq
H_0 ~=~ v_F ~\sum_i ~\sum_\al ~\int_{- \La}^\La ~
\frac{dk}{2\pi} ~k~ c_{i \al k}^\dg c_{i \al k} ~.
\label{h0}
\eeq

If the impurity spin is coupled to the electrons at the junction, that part 
of the Hamiltonian is given by
\beq
H_{\rm spin} ~=~ \sum_{\al ,\be} ~J ~{\vec S} \cdot \Psi_\al^\dg
(x=0) ~\frac{{\vec \si}_{\al\be}}{2} \Psi_\be (x=0)~,
\label{hspin1}
\eeq
where ${\vec \si}$ denotes the Pauli matrices.
(For simplicity, we will assume an isotropic spin coupling $J_x = J_y = J_z$).
Eq. (\ref{hspin1}) can be written in terms of second quantized operators as 
\bea
H_{\rm spin} &=& \sum_{i,j} \sum_{\al ,\be} ~\int_{-\La}^\La ~\int_{-\La}^\La ~
\frac{dk_1}{2\pi} ~\frac{dk_2}{2\pi} \non \\
& & ~~~~~~~~~~~~~ J_{ij} ~{\vec S} \cdot ~c_{i\al k_1}^\dg ~
\frac{{\vec \si}_{\al \be}}{2} ~c_{j\be k_2} ~, 
\label{hspin2}
\eea
where $J_{ij} = J (1 + \sum_l S_{li}^*) (1 + \sum_m S_{mj})$ is a Hermitian
matrix. In general, however, the impurity spin may also be coupled to the 
electrons at other sites which are slightly away from the junction; for 
instance, this may be true if the model is defined on a lattice at the 
microscopic scale as we will see in Sec. V. It is therefore convenient to 
take $J_{ij}$ to be an arbitrary Hermitian matrix which is not necessarily 
related to the entries of the $S$-matrix in any simple way.

Next, let us consider density-density interactions between the electrons in
each wire of the form (we will drop the wire index $i$ for the moment)
\beq
H_{\rm int} ~=~ \frac{1}{2} ~\int \int ~dx ~dy ~\rho (x) ~U(x-y) ~\rho (y) ~,
\label{hint1}
\eeq
where the density $\rho$ is given in terms of the second quantized electron 
field $\Psi_\al (x)$ as $\rho = \Psi^\dg_\up \Psi_\up +\Psi^\dg_\dn 
\Psi_\dn$. As mentioned earlier for the wave-functions, the electron field 
can also be written in terms of outgoing and incoming fields as
\beq
\Psi_\al (x) ~=~\Psi_{O\al} (x) ~e^{ik_F x} ~+~\Psi_{I\al} (x) ~e^{-ik_F x} ~.
\label{psirl}
\eeq
If the range of the interaction $U(x)$ is short (of the order of the Fermi 
wavelength $2\pi /k_F$), such as that of a screened Coulomb repulsion, 
the Hamiltonian in (\ref{hint1}) can be written as
\bea
& & H_{\rm int} = \non \\
& & \int dx \sum_{\al ,\be} [g_1 \Psi_{O\al}^\dg \Psi_{I\be}^\dg \Psi_{O\be} 
\Psi_{I\al}+g_2 \Psi_{O\al}^\dg \Psi_{I\be}^\dg \Psi_{I\be} \Psi_{O\al} \non \\
& & ~~~~~~+ \frac{1}{2} g_4 ( \Psi_{O\al}^\dg \Psi_{O\be}^\dg \Psi_{O\be} 
\Psi_{O\al} + \Psi_{I\al}^\dg \Psi_{I\be}^\dg \Psi_{I\be} \Psi_{I\al} )], 
\non \\
& &
\label{hint2}
\eea
where
\bea
g_1 ~&=&~ {\tilde U} (2k_F) ~, \non \\
{\rm and} ~~~g_2 ~&=&~ g_4 ~=~ {\tilde U} (0) ~.
\eea
For repulsive and attractive interactions, $g_2 > 0$ and $< 0$ respectively.
(We have ignored umklapp scattering terms here; they only arise if the 
model is defined on a lattice and we are at half-filling).

\section{\bf III. The renormalization group equations}

It is known that the interaction parameters $g_1$, $g_2$ and $g_4$ satisfy 
some RG equations \cite{solyom}; the solutions of the lowest order RG 
equations are given by \cite{yue}
\bea
g_1 (L) &=& \frac{{\tilde U} (2k_F)}{1 ~+~ \frac{{\tilde U} (2k_F)}{\pi v_F}
\ln L} ~, \non \\
g_2 (L) &=& {\tilde U} (0) ~-~ \frac{1}{2} ~{\tilde U} (2k_F) ~+~ \frac{1}{2}~
\frac{{\tilde U} (2k_F)}{1 ~+~ \frac{{\tilde U} (2k_F)}{\pi v_F} \ln L} ~,
\non \\
g_4 (L) &=& {\tilde U} (0) ~,
\label{g124}
\eea
where $L$ denotes the length scale.

In general, the couplings $g_1$, $g_2$ and $g_4$ can have different values on 
different wires; hence we have to add a subscript $i$ to them. For weak 
interactions, i.e., when $g_{1i}$, $g_{2i}$ and $g_{4i}$ are all much less than
$2 \pi v_F$, we can derive the RG equations 
for the $S$-matrix at the junction \cite{yue,lal1}. Let us define a parameter
\beq
\al_i ~=~ \frac{g_{2i} ~-~ 2 ~g_{1i}}{2\pi v_F} ~,
\eeq
which is a function of length scale due to Eqs. (\ref{g124}), and a diagonal 
matrix $M$ whose entries are given by
\beq
M_{ii} ~=~ \frac{1}{2} ~\al_i r_{ii} ~.
\eeq
Then the RG equations can be written in the matrix form
\beq
\frac{dS}{d \ln L} ~=~ M ~-~ S M^\dg S ~.
\label{rgs}
\eeq
The FPs of this equation are given by the condition $M = S M^\dg S$.

\begin{figure}[htb]
\begin{center}
\epsfig{figure=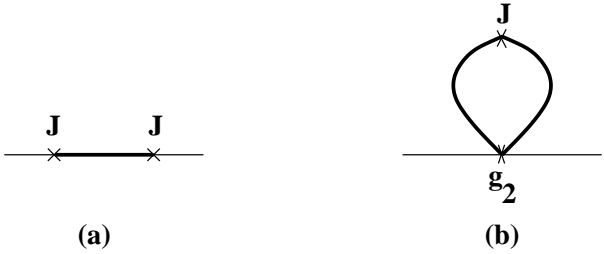,width=8cm}
\end{center}
\caption{Pictures of the terms which contribute to the renormalization of the
Kondo coupling matrix $J$ to order $J^2$ and $g_2 J$ respectively; $g_2$ 
denotes the coefficient of the electron-electron interaction. Thin lines and 
thick lines denote low energy and high energy electrons respectively.}
\end{figure}

We use the technique of `poor man's RG' \cite{anderson,noz1} to derive the 
renormalization of the $S$-matrix and the Kondo coupling matrix $J_{ij}$.
Briefly, this involves using the second order perturbation expression for
the low energy effective Hamiltonian,
\beq
H_{\rm eff} ~=~ \sum_h ~\frac{|l_2 > < l_2 | H_{\rm pert} | h > < h | 
H_{\rm pert} | l_1> < l_1 |}{E_l - E_h} ~,
\label{heff1}
\eeq
where the perturbation $H_{\rm pert}$ is given by the sum of $H_{\rm spin}$
and $H_{\rm int}$ in Eqs. (\ref{hspin2}) and (\ref{hint2}), 
$l_1$ and $l_2$ denote two energy states, and $h$ denotes high energy 
states. We now restrict the sum over $h$ in Eq. (\ref{heff1}) to run over 
states for which the energy difference $E_h - E_l$ lies within an energy 
shell $E$ and $E+dE$; we have assumed that the difference between 
different low energy states is much smaller than $E$, so that we can simply 
write $E_{l_1} = E_l$ in the denominator of the above equation. We then see 
that the change in the effective Hamiltonian $dH_{\rm eff}$ is proportional 
to $dE /E$ which is equal to $- d \ln L$, where the length scale $L$ is 
inversely related to the energy scale $E$. In this way, we get an RG
equation for the derivatives with respect to $\ln L$ of various parameters 
appearing in the low energy Hamiltonian.

Using this method, we find that the Kondo couplings $J_{ij}$ do not contribute
to the renormalization of the $S$-matrix in Eq. (\ref{rgs}) up to second order
in $J_{ij}$. (This is not true beyond second order; however, we will only work
to second order here assuming that the $J_{ij}$ are small). On the other hand,
the $S$-matrix does contribute to the renormalization of the $J_{ij}$ 
through the interaction Hamiltonian in Eq. (\ref{hint2}); this is because the 
relation between the outgoing field on wire $i$ (i.e., $\Psi_{O i \al}$) 
and the operators $c_{j \al}$ involves the $S$-matrix. For instance, the
terms involving $g_{2i}$ in Eq. (\ref{hint2}) take the form
\bea
& & \sum_{i,j,l} \sum_{\al ,\be} ~\int_{-\La}^\La ~\int_{-\La}^\La ~
\int_{-\La}^\La ~\int_{-\La}^\La ~\frac{dk_1}{2\pi} ~\frac{dk_2}{2\pi} ~
\frac{dk_3}{2\pi} ~\frac{dk_4}{2\pi} \non \\
& & \times ~\pi ~\de (k_1 - k_2 + k_3 - k_4) ~g_{2i} \non \\
& & \times ~S^*_{ij} c_{j\al k_1}^\dg ~c_{i\be k_2}^\dg ~c_{i\be k_3}~ S_{il}
c_{l\al k_4} ~,
\label{g2term}
\eea
where we have used the identity
\bea
& & \int_0^\infty ~dx ~e^{(-ik_1 + ik_2 - ik_3 + ik_4 -\ep)x} \non \\
& & = ~-~ \frac{i}{k_1 - k_2 + k_3 - k_4 - i \ep} \non \\
& & = ~-~ i ~{\rm P} \Bigl( \frac{1}{k_1 - k_2 + k_3 - k_4} \Bigr) \non \\
& & ~~~~+~ \pi ~\de (k_1 - k_2 + k_3 - k_4) ~,
\eea
with $\ep$ being an infinitesimal positive number. [In Eq. (\ref{g2term}),
we have kept only the $\de$-function term and have dropped the principal 
part term since the latter can be either positive or negative, and its 
contribution vanishes when one integrates over the variables $k_i$.] Note that
the terms involving $g_2$ in Eq. (\ref{g2term}) (as well as those involving
$g_1$ and $g_4$ in Eq. (\ref{hint2})) conserve momentum while the Kondo 
coupling terms in Eq. (\ref{hspin2}) do not.

We will omit the details of the RG calculations here apart from making
a few comments below. We find that
\bea
& & \frac{dJ_{ij}}{d \ln L} \non \\
& & =~ \frac{1}{2\pi v_F} ~[~ \sum_k ~J_{ik} J_{kj} \non \\
& & ~~~~+~ \frac{1}{2} ~g_{2i} ~S_{ij} ~\sum_k ~J_{ik} S^*_{ik} ~+~ 
\frac{1}{2} ~g_{2j} ~S^*_{ji} ~\sum_k ~ J_{kj} S_{jk} \non \\
& & ~~~~-~ \frac{1}{2} \sum_k ~(g_{2k} - 2 g_{1k}) ~(J_{ik} S^*_{kk} S_{kj} + 
S^*_{ki} S_{kk} J_{kj} ) ~]~, \non \\
& &
\label{rgj}
\eea
where $S_{ij}$ is the $S$-matrix at the length scale $L$. Eq. (\ref{rgj}) is 
the key result of this paper. Note that it maintains the hermiticity of
the matrix $J_{ij}$. Eq. (\ref{rgj}) always has a trivial FP at $J_{ij} =0$.

Let us briefly comment on the origin of the various 
terms on the right hand side of Eq. (\ref{rgj}). The first and second lines 
arise from Figs. 1 (a) and (b) respectively. (The terms 
of order $J^2$ in the first line have been derived in Ref. \cite{rosch}). 
The parameters $g_{1i}$ and $g_{4i}$ do not appear in the second
line of Eq. (\ref{rgj}) since the terms which are proportional
to these parameters either do not appear in the numerator of Eq. (\ref{heff1})
because they are not allowed by momentum conservation, or they appear in Eq.
(\ref{heff1}) but their contribution vanishes because the Pauli matrices are 
traceless. Finally, the third line of Eq. (\ref{rgj}) arises as follows.
In Ref. \cite{lal1}, the RG equation for the $S$-matrix was derived.
This was based on the idea that due to reflections at the junction
(these arise from the diagonal elements of the $S$-matrix which are the
reflection amplitudes), there are
Friedel oscillations in the density of the electrons; the amplitudes
of these oscillations are proportional to $S_{kk}$ and $S^*_{kk}$ in wire $k$.
We now treat the interactions in the Hartree-Fock approximation \cite{lal1};
this results in reflections from the Friedel oscillations with a
strength proportional to $g_{2k} - 2 g_{1k}$ in wire $k$.
Now, an electron going from wire $j$ to $i$ can either
(i) first go from wire $j$ to wire $k$ with a transmission amplitude 
$S_{kj}$, scatter from the Friedel oscillations in wire $k$ with an amplitude 
$(g_{2k} - 2 g_{1k}) S^*_{kk}$, and finally scatter off the impurity spin 
from wire $k$ to wire $i$ with amplitude $J_{ik}$, or
(ii) first scatter off the impurity spin from wire $j$ to wire $k$ with 
amplitude $J_{kj}$, scatter from the Friedel oscillations in wire $k$ with an 
amplitude $(g_{2k} - 2 g_{1k}) S_{kk}$, and finally scatter from wire $k$
to wire $i$ with a transmission amplitude $S^*_{ki}$.
These two processes give rise to the third line of Eq. (\ref{rgj}).

It is interesting to observe that Eq. (\ref{rgj}) remains invariant if 
we transform $S_{ij} \to e^{i\phi_i} S_{ij}$, where the $\phi_i$ can be 
arbitrary real numbers. According to Eq. (\ref{wavefn}), this corresponds to 
the freedom of redefining the phases of the outgoing waves by different 
amounts on different wires. 

\section{\bf IV. Analysis of the renormalization group equations}
 
To simplify our analysis, we will make two assumptions.

\noi (i) The couplings $g_{1i}$ and $g_{2i}$ have the same value on all the 
wires, and therefore the subscript $i$ on $g_1$ and $g_2$ can be dropped. 

\noi (ii) The $S$-matrix is at a FP of Eq. (\ref{rgs}), so that
$S$ does not flow with the length scale.

We will now consider several possibilities for the $S$-matrix, and will study
the RG flows and FPs of the Kondo couplings $J_{ij}$ in each case. The
different possibilities can be realized in terms of quantum wires and 
quantum dots containing the impurity spin as shown in Fig. 2.

\begin{figure}[htb]
\begin{center}
\epsfig{figure=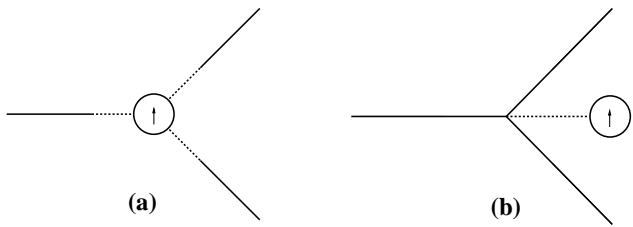,width=8.5cm}
\end{center}
\caption{Schematic pictures of the system of wires (shown by solid lines), 
an impurity spin (shown inside a circle), and the coupling between the
spin and the wires (dotted lines). Figures (a) and (b) show the cases of 
disconnected and Griffiths $S$-matrices respectively.}
\end{figure}

\subsection{\bf A. $N$ disconnected wires}

The $S$-matrix for $N$ disconnected wires is given by the $N \times N$ 
identity matrix (up to phases). (We will assume that $N \ge 2$). A picture
of the system is indicated in Fig. 2 (a); the wires are disconnected
from each other, and the end of each wire is coupled to the impurity spin. 
A more microscopic description of the system will be discussed in Sec. V.

Let us consider a highly symmetric form of the Kondo coupling matrix in which 
all the diagonal entries are equal to $J_1$ and all the off-diagonal entries 
are equal to $J_2$, with both $J_1$ and $J_2$ being real. (In the language
of the three-dimensional $N$-channel Kondo problem, $J_2$ denotes coupling
between different channels). Since the $S$-matrix
is also symmetric under the exchange of any two of the $N$ indices, such a 
symmetric form of the Kondo matrix will remain intact during the course of the
RG flow. In other words, it is natural for us to choose the $J$ matrix to have
the same symmetry as the $S$-matrix, since that symmetry is preserved under 
the RG flow. Eq. (\ref{rgj}) gives the two-parameter RG equations
\bea
\frac{dJ_1}{d \ln L} &=& \frac{1}{2\pi v_F} ~[ J_1^2 + (N-1) J_2^2 + 2 g_1
J_1 ] ~, \non \\
\frac{dJ_2}{d \ln L} &=& \frac{1}{2\pi v_F} ~[ 2 J_1 J_2 + (N-2) J_2^2 -
(g_2 - 2 g_1) J_2 ]~. \non \\
& &
\label{disc}
\eea
(For $N=2$ and $g_1 = 0$, Eq. (\ref{disc}) agrees with the results in Ref. 
\cite{fabrizio}).

Since $g_1 (L=\infty) = 0$, Eq. (\ref{disc}) has only one FP at finite values 
of $(J_1, J_2)$, namely, the trivial FP at $(0,0)$. We then carry out a linear
stability analysis around this FP. [Given a RG equation of the form
$dX/d \ln L = a X$, we will say that the FP at $X=0$ is stable if $a< 0$, 
unstable if $a> 0$, and marginal if $a=0$. In the marginal case, we look at 
the next order term; if $dX/d \ln L = bX^2$ and $b>0$, we say that the FP at 
$X=0$ is stable on the $x<0$ side and unstable on the $x>0$ side.] 
If $\nu \equiv g_2 (L=\infty)/(2 \pi v_F) > 0$
(repulsive interactions), the stability analysis shows that the trivial FP
is stable to small perturbations in $J_2$. For small perturbations in $J_1$, 
this FP is marginal; a second order analysis shows that it is stable if $J_1 
< 0$ and unstable if $J_1 > 0$, i.e., it is the usual {\it ferromagnetic} 
FP which is found for Fermi liquid leads. However, the approach to 
the FP is quite different when the leads are TLLs. At large length 
scales, the FP is approached as $J_1 \sim - 1/\ln L$ and $J_2 \sim 1 /L^\nu$.
{}From this, we can deduce the behavior at very low temperatures, namely,
\beq
J_1 ~\sim~ -~ 1/\ln (T_K /T) ~, ~~{\rm and} ~~~ J_2 ~\sim~ (T/T_K)^\nu ~.
\label{lowt1}
\eeq
where we have introduced the Kondo temperature $T_K$. (This is given as usual
by $T_K \sim \Lambda e^{-2\pi v_F /J}$, where $\Lambda$ is an energy cut-off 
of the order of the Fermi energy $E_F$, $J$ is the value of a typical Kondo
coupling at the microscopic length scale as explained after Eq. (\ref{asymp}),
and $1/(2\pi v_F)$ is the density of states at $E_F$). The form in Eq. 
(\ref{lowt1}) is in contrast to the behavior of $J_2$ for Fermi liquid leads,
i.e., for $g_1 = g_2 = 0$. In that case, Eq. (\ref{disc}) can be solved
exactly in terms of the linear combinations $J_1 - J_2$ and
$J_1 + (N-1) J_2$; we again find a FP at $(J_1,J_2) = (0,0)$, with 
\beq
J_1 ~\sim~ -~ 1/\ln (T_K /T) ~, ~~{\rm and} ~~~ J_2 ~\sim~ 1/\ln (T_K /T)^2 ~.
\label{lowt2}
\eeq
Note that $J_2$ approaches zero faster than $J_1$ for both Fermi liquid leads 
and TLL leads; but for the latter case, it goes to zero much faster, i.e., as 
a power of $T$. 

Eq. (\ref{lowt2}) is valid provided that neither $J_1 - J_2$ nor $J_1 + 
(N-1) J_2$ is {\it exactly} equal to zero; if one of them is exactly zero and 
the other is not, then both $J_1$ and $J_2$ go as $1/\ln (T_K /T)$. However, 
having one of the two combinations exactly equal to zero requires a special 
tuning in a microscopic model, as we will see in Sec. V. In general, 
therefore, the powers of $1/ \ln (T_K /T)$ in $J_1$ and $J_2$ are different;
this does not seem to have been noted in the earlier literature. 

\begin{figure}[htb]
\begin{center}
\epsfig{figure=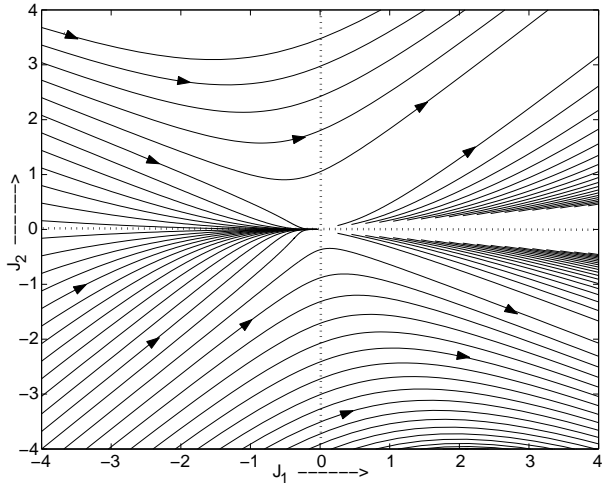,width=8cm}
\end{center}
\caption{RG flows of the Kondo couplings for three disconnected wires, with
${\tilde U} (0) = {\tilde U} (2k_F) = 0.2 (2 \pi v_F)$.}
\end{figure}

Figure 3 shows a picture of the RG flows for three wires for ${\tilde U} (0)
= {\tilde U} (2k_F) = 0.2 (2 \pi v_F)$. [This gives a value of $\nu$ which is 
comparable to what is found in several experimental systems (see \cite{lal2} 
and references therein). In all the pictures of RG flows, the values of 
$J_{ij}$ are shown in units of $2\pi v_F$.] We see that the RG flows take a 
large range of initial conditions to the FP at $(0,0)$. For all other initial 
conditions, we see that there are two directions along which the Kondo 
couplings flow to large values; these are given by $J_2 /J_1 = 1$ and 
$J_2 /J_1 = - 1/(N-1)$ (with $N=3$). 
[On a cautionary note, we should remember that the RG equations studied here 
are only valid at the lowest order in $J_{ij}$ and $g_2$, i.e., for the case 
of weak repulsion (or attraction) and small Kondo couplings.]

The fact that the Kondo couplings flow to large values along two particular
directions can be understood as follows. For values of $J_1$ and 
$J_2$ much larger than $g_1$ and $g_2$, one can ignore the terms of order 
$g_1$ and $g_2$ in Eq. (\ref{disc}). One then obtains the decoupled equations
\bea
\frac{d~[J_1 - J_2]}{d \ln L} &\simeq& \frac{1}{2\pi v_F} ~(J_1 - J_2)^2 ~, 
\non \\
\frac{d~[J_1 + (N-1)J_2]}{d \ln L} &\simeq& \frac{1}{2\pi v_F} ~(J_1 + (N-1) 
J_2)^2 ~. \non \\
& &
\label{asymp}
\eea
{}From these equations one can deduce that the couplings can flow to 
large values in one of two ways, depending on the initial conditions. 
Either $J_1 + (N-1) J_2$ goes to $\infty$ much faster than $J_1 - J_2$ 
(this is what happens in the first quadrant of the figures in
Figs. 3 and 4), or $J_1 - J_2$ goes to $\infty$ much faster than $J_1 + 
(N-1) J_2$ (this happens in the fourth quadrant of the figures in 
Figs. 3 and 4). A third possibility is that $J_2$ remains exactly equal to 
zero while $J_1 \to \infty$; however, this can only happen if one begins with 
$J_2$ exactly equal to zero. (This also seems to happen if the interactions
are strong enough as we will discuss in Sec. VII). We will provide a physical 
interpretation of the first two possibilities in Sec. V.

{}Eq. (\ref{asymp}) has the form $dJ /d \ln L = J^2 /(2\pi v_F)$. If 
$J(d)$ denotes the value of $J$ at a microscopic length $d$, and $J(d) \ll
2 \pi v_F$, then it becomes of order 1 at a length scale $L_0$ given by
$L_0 /d \sim e^{2\pi v_F /J(d)}$; the corresponding temperature is
given by $T_K \sim \Lambda e^{-2\pi v_F /J(d)}$.

Finally, note that the special case with $J_2 = 0$ and $g_1 = g_2 = 0$ is 
equivalent to the Kondo problem in three dimensions with $N$ channels and no 
coupling between channels \cite{kondo}. In the three-dimensional case, the RG 
equation has been derived to fifth order in the Kondo coupling \cite{gan2}. 
This reveals a stable FP at a finite value of the coupling 
\beq
J_1 ~=~ \frac{4\pi v_F}{N} ~,
\label{finitej}
\eeq
where $1/(2\pi v_F)$ is the density of states at the Fermi energy. Thus
the couplings $J_{ij}$ need not really flow to infinity as Fig. 3 would 
suggest; one may find strong coupling FPs lying at values of order $2\pi v_F$ 
if one takes into account terms of higher order in the RG equations. In Sec. 
VII, we do find a strong coupling FP for sufficiently strong inter-electron 
interactions.

Although we have discussed the case of completely disconnected wires here, the
results do not change significantly if we allow a small spin-independent 
tunneling amplitude of the form
\beq
H_{\rm tun} ~=~ \tau ~\sum_{i \ne j} ~\sum_\al ~\Psi_{i,\al}^\dg (x_i=0) ~
\Psi_{j,\al} (x_j=0)~.
\label{tun}
\eeq
This is equivalent to changing the $S$-matrix slightly away from the identity 
matrix. Using the RG equation in (\ref{rgs}), we find that the parameter
$\tau$ satisfies the RG equation 
\beq
\frac{d\tau}{d \ln L} ~=~ - ~\frac{1}{2\pi v_F} ~(g_2 ~-~ 2 g_1) ~\tau ~.
\label{tau}
\eeq
This has the same form as the interaction dependent terms in the RG equation
for $J_2$ in (\ref{disc}). Hence, $\tau$ also scales at low temperatures as 
$T^\nu$ just like $J_2$ in Eq. (\ref{lowt1}). Thus the contributions of both
$\tau$ and $J_2$ to the conductance go as $(T/T_K)^{2\nu}$.

Here and subsequently we have not discussed the case of attractive 
interactions ($g_2 < 0$).
The stability analysis can easily be suitably modified in that case; some
of the directions for the RG flows may become stable and others may
become unstable if the sign of $g_2$ is reversed.

\subsection{\bf B. Griffiths $S$-matrix for $N$ wires}

This is the case in which all the $N$ wires are connected to each other and 
there is maximal transmission, subject to the constraint that there is 
complete symmetry between the $N$ wires. (We will again assume that $N \ge 
2$.) A picture of the system is indicated in Fig. 2 (b); the wires are 
connected to each other at a junction, and the junction is also coupled to 
the impurity spin. A more microscopic description of the junction will be 
discussed in Sec. V.

The maximally transmitting completely symmetric 
$S$-matrix is also called the Griffiths $S$-matrix and has all the diagonal 
entries equal to $-1 + 2/N$ and all the off-diagonal entries equal to $2/N$.
Since here, too, the $S$-matrix is fully symmetric in the $N$ wires, we
again consider the highly symmetric form of the Kondo coupling matrix as 
in the previous subsection, with real parameters $J_1$ and $J_2$ as the 
diagonal and off-diagonal entries respectively. Eq. (\ref{rgj}) then gives
\bea
\frac{dJ_1}{d \ln L} &=& \frac{1}{2\pi v_F} ~[J_1^2 + (N-1) J_2^2 ~+~ 
2g_1 ~(1 - \frac{2}{N})^2 ~J_1 \non \\
& & ~~~~~~-~ 4g_1 (1 - \frac{2}{N})~ (1 - \frac{1}{N}) ~J_2 ~], \non \\
\frac{dJ_2}{d \ln L} &=& \frac{1}{2\pi v_F} ~[ 2 J_1 J_2 +(N-2) J_2^2 ~-~ 
\frac{4g_1}{N} ~(1 - \frac{2}{N}) J_1 \non \\
& & ~~~~~~+~ (g_2 - 2 g_1 (1 - \frac{2}{N})^2) ~J_2 ~].
\label{grif}
\eea
(For $N=2$, i.e., a full line with an impurity spin coupled to one point 
on the line, Eq. (\ref{grif}) agrees with the equations derived in Ref. 
\cite{furusaki}).

The only FP of Eq. (\ref{grif}) is again the trivial FP at the origin. A 
linear stability analysis shows that this FP is unstable in one direction 
($J_2$) and marginal in the other ($J_1$) for $g_2 (L=\infty) >0$. 

Figure 4 shows a picture of the RG flows for three wires for ${\tilde U} (0)
= {\tilde U} (2k_F) = 0.2 (2 \pi v_F)$. We see that there is no stable FP at 
finite values of the couplings. The couplings flow to large values along one 
of the two directions $J_2 /J_1 = 1$ and $J_2 /J_1 = -1 /(N-1)$. The reason 
for this is the same as that explained around Eq. (\ref{asymp}) since the RG 
equations in (\ref{disc}) and (\ref{grif}) have the same form for large 
values of $J_1$ and $J_2$.

\begin{figure}[htb]
\begin{center}
\epsfig{figure=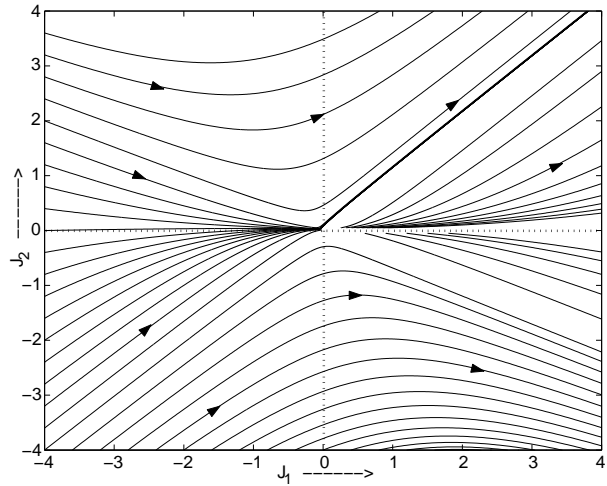,width=8cm}
\end{center}
\caption{RG flows of the Kondo couplings for the Griffiths $S$-matrix for 
three wires, with ${\tilde U} (0) = {\tilde U} (2k_F) = 0.2 (2 \pi v_F)$.}
\end{figure}

\subsection{\bf C. Chiral $S$-matrix for three wires}

We choose a chiral $S$-matrix of the form
\bea
S &=& \left( \begin{array}{ccc}
0 & 0 & \gamma \\
\gamma & 0 & 0 \\
0 & \gamma & 0 \end{array} \right) ~,
\label{chirs}
\eea
where $\gamma$ is a complex number satisfying $|\gamma | =1$. (We will see a 
physical realization of this form in Sec. V. Alternatively, we could have 
considered an $S$-matrix which is the transpose of the one given above).

\begin{figure}[htb]
\begin{center}
\epsfig{figure=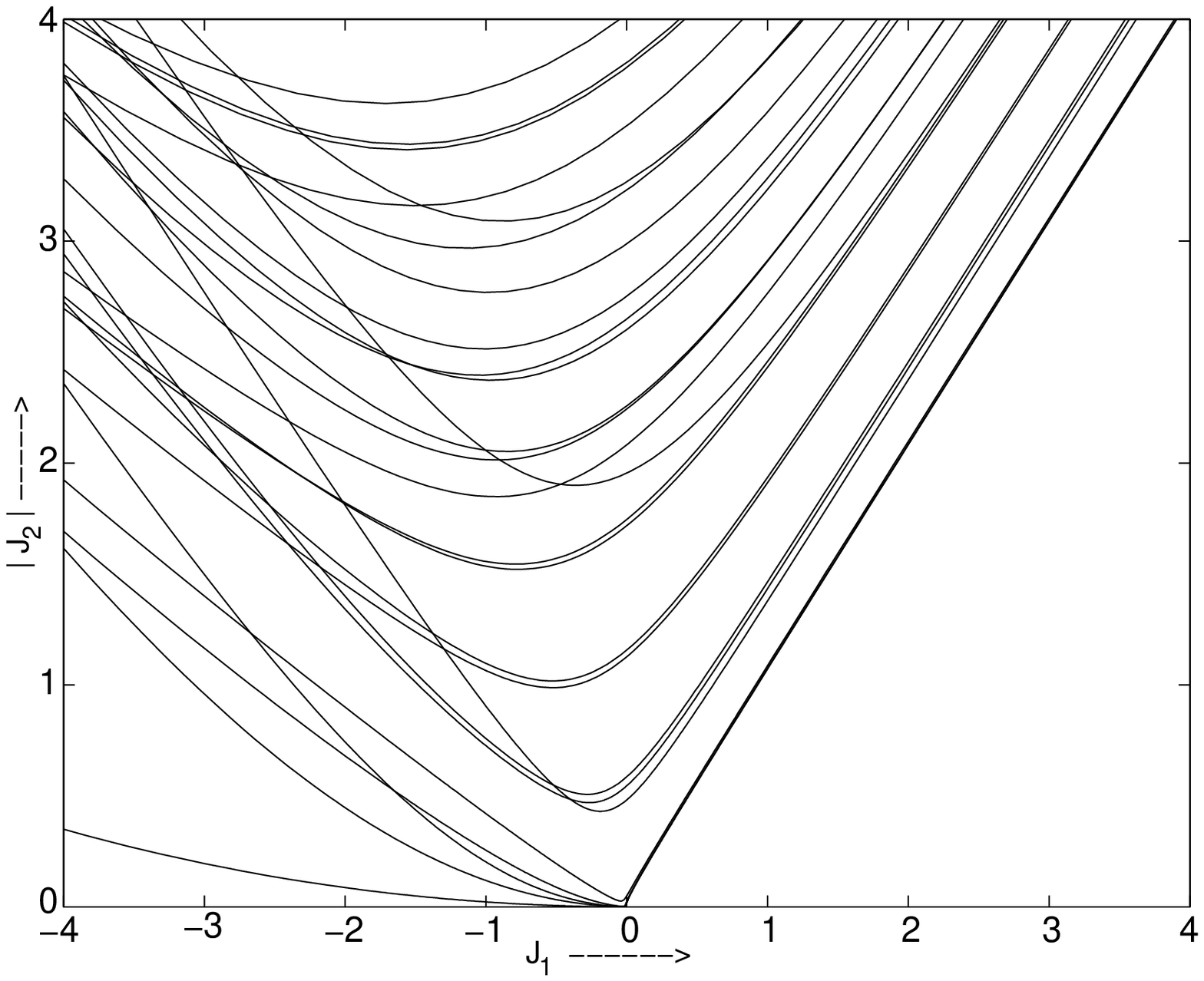,width=8cm}
\epsfig{figure=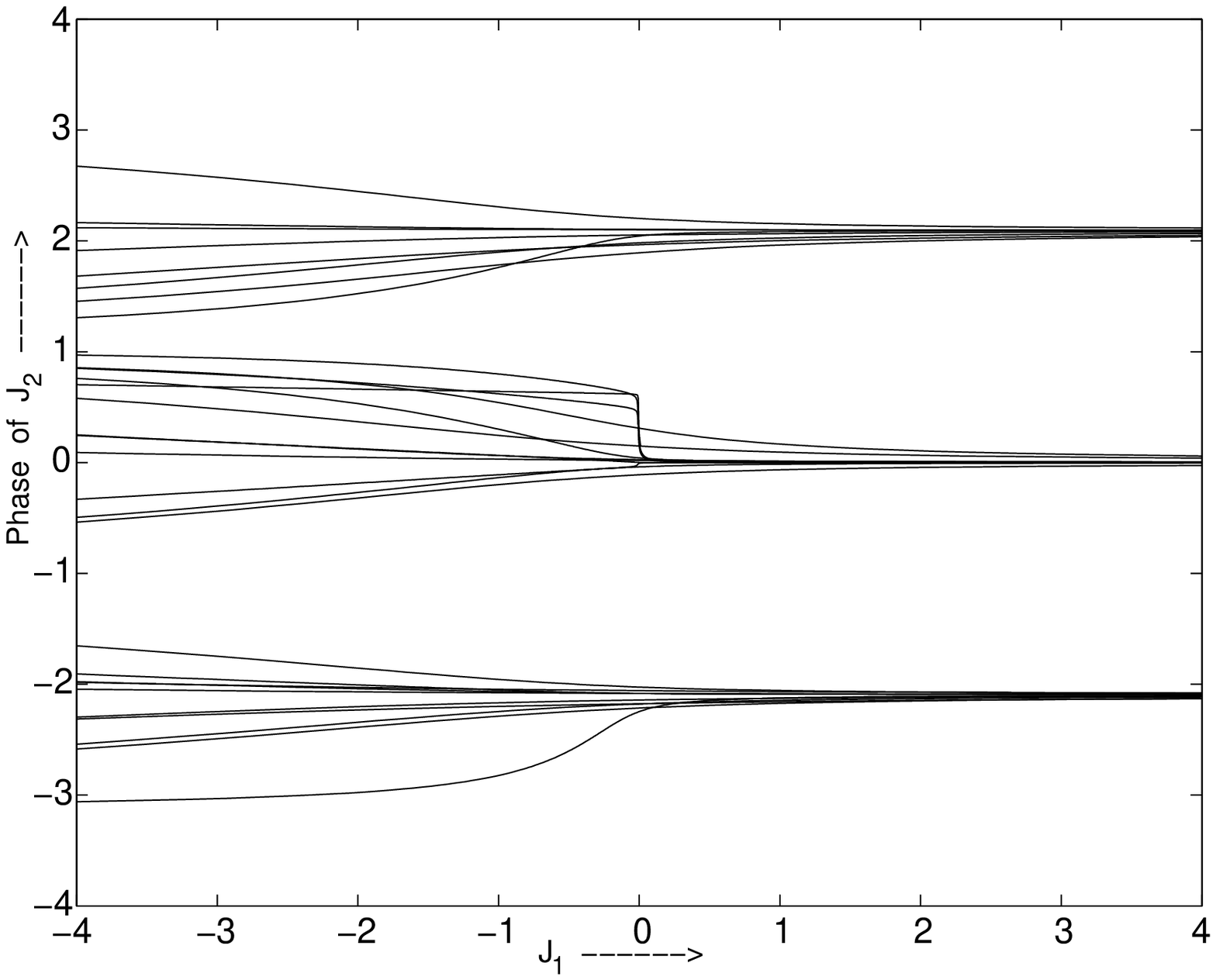,width=8cm}
\end{center}
\caption{RG flows for the chiral $S$-matrix for three wires, with
${\tilde U} (0) = {\tilde U} (2k_F) = 0.2 (2 \pi v_F)$. The upper and lower 
figures show the magnitude and phase respectively of $J_2$.}
\end{figure}

Let us consider a Kondo coupling matrix of the form
\bea
J &=& \left( \begin{array}{ccc}
J_1 & J_2 & J^*_2 \\
J^*_2 & J_1 & J_2 \\
J_2 & J^*_2 & J_1 \end{array} \right) ~,
\label{konchi}
\eea
where $J_1$ is real but $J_2$ can be complex.

Then Eq. (\ref{rgj}) gives
\bea
\frac{dJ_1}{d \ln L} &=& \frac{1}{2\pi v_F} ~[~ J_1^2 + 2 |J_2|^2 ~] ~, \non \\
\frac{dJ_2}{d \ln L} &=& \frac{1}{2\pi v_F} ~[~ 2 J_1 J_2 + (J^*_2)^2 ~+~ 
\frac{1}{2} ~g_2 J_2 ~] ~.
\label{chir1}
\eea
[Note that the above equations remain invariant under the transformation
$J_2 \to J_2 e^{i 2 \pi /3}$ or $J_2 e^{-i 2 \pi /3}$. We will see in
Sec. V. C that a lattice realization of the chiral $S$-matrix has the
same symmetry.]

One can again show that the only FP of Eq. (\ref{chir1}) is the trivial FP at
the origin. A linear stability analysis shows that the trivial FP is unstable 
in one direction ($J_2$) and marginal in the other ($J_1$) for $g_2 (L=\infty)
>0$. Figure 5 shows a picture of the RG flows for three wires for ${\tilde U}
(0) = {\tilde U} (2k_F) = 0. 2 (2 \pi v_F)$. The upper and lower figures show
the way in which the magnitude and phase of $J_2$ evolve. We see that there 
is no stable FP at finite values of the couplings. The phase of $J_2$ flows 
towards one of the three values, 0 or $\pm 2 \pi /3$; this is consistent with 
the symmetry of $J_2$ pointed out after Eq. (\ref{chir1}). Further, $J_1$ and 
the magnitude of $J_2$ flow in such a way that $J_1 + 2 |J_2|$ grows much
faster than $J_1 - |J_2|$. We can understand these observations as follows.

For values of $J_1$ and $J_2$ much larger than $g_2$, one can ignore
the term of order $g_2$ in Eq. (\ref{chir1}). If we write $J_2 = |J_2| e^{i
\phi}$, we find that 
\bea
\frac{d\phi}{d \ln L} &\simeq& -~ \frac{1}{2\pi v_F} ~|J_2| ~\sin (3 \phi) ~,
\non \\
\frac{d|J_2|}{d \ln L} &\simeq& \frac{1}{2\pi v_F} ~[~ 2 J_1 |J_2| + |J_2|^2 ~
\cos (3 \phi) ~] ~.
\label{chir2}
\eea
The first equation in (\ref{chir2}) shows that $\phi = 0, \pm \pi /3, \pm 2 
\pi /3$ and $\pi$ are fixed points; however, since $|J_2|$ flows to $\infty$ 
under RG, only the values $\phi = 0$ and $\pm 2 \pi /3$ are stable.
Substituting this fact that $\cos (3 \phi) \to 1$
in the second equation in (\ref{chir2}), and combining it with the first 
equation in (\ref{chir1}), we obtain the decoupled equations
\bea
\frac{d~[J_1 - |J_2|]}{d \ln L} &\simeq& \frac{1}{2\pi v_F} ~(J_1 - |J_2|)^2 ~,
\non \\
\frac{d~[J_1 + 2|J_2|]}{d \ln L} &\simeq& \frac{1}{2\pi v_F} ~(J_1 + 2 
|J_2|)^2 ~. 
\eea
{}From this we deduce that $J_1 + 2 |J_2|$ must flow to $\infty$ much faster
than $J_1 - |J_2|$ since $J_1 + 2 |J_2| > J_1 - |J_2|$ to begin with. Note 
that unlike the disconnected and Griffiths cases, where $J_1$ and $J_2$ 
flow to large values in two possible ways (with $|J_2|/J_1 \to 1$ and 
$-1/(N-1)$ respectively), in the chiral case, $J_1$ and $J_2$ flow to 
large values in only one way, along the direction $|J_2|/J_1 = 1$.

\section{\bf V. Interpretation in terms of lattice models}

We will now see how the different $S$-matrices and RG flows discussed in Sec. 
IV can be interpreted in terms of lattice models \cite{furusaki}. This will
provide us with physical interpretations of the various kinds of RG flows and
FPs. We will concentrate on what the lattice models imply about the structure 
of the region near the junction, rather than the form of the interactions 
between the electrons in the bulk of the wires which has already been 
discussed in Sec. II. (The interactions can be introduced in the lattice model
by, for instance, writing a Hubbard term at each site). We will again discuss 
three different cases here.
(The models shown in Fig. 6 and discussed below in detail can be thought of
as providing a microscopic picture of the systems shown in Fig. 2).

\begin{figure}[htb]
\begin{center}
\epsfig{figure=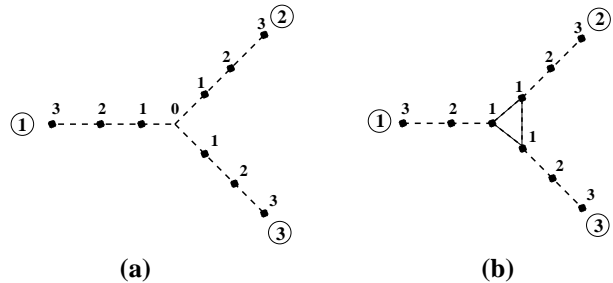,width=8.5cm}
\end{center}
\caption{Lattice models for some of the $S$-matrices for three wires.
(a) can be a model for the disconnected and Griffiths $S$-matrices,
while (b) can be a model for the chiral $S$-matrix.}
\end{figure}

\subsection{\bf A. $N$ disconnected wires}

This system can be realized by a lattice of the form shown in Fig. 6 (a). $N$ 
wires meet at a junction which is labeled by the site number 0; all the other 
sites are labeled as $n=1, 2, \cdots$, with $n$ increasing as one goes away 
from the junction. (The lattice spacing will be set equal to one). We take 
the Hamiltonian to be of the tight-binding form, with a hopping amplitude 
equal to $-t$ on all the bonds (where $t$ is real), except for the bonds 
which connect the sites labeled as $n=1$ on each wire to the junction site; 
we set those hopping amplitudes equal to zero. This is equivalent to removing 
the junction site from the system; we will therefore not consider that site 
any further in this subsection. We then obtain a system of disconnected
wires with an $S$-matrix which is equal to $-1$ times the identity matrix.
To show this, we consider a wave which is incoming on wire $i$ with a
wave number $k$, where $0 < k < \pi$. We then find that the corresponding 
eigenstate of the Hamiltonian has an energy equal to $E_k = - 2 t \cos k$,
and a wave function given by
\bea
\psi_{ik} (n) &=& e^{-ikn} ~-~ e^{ikn} ~~{\rm for} ~~
n=1,2,\cdots ~~{\rm on ~wire~} i ~, \non \\
&=& 0 ~~{\rm at ~the ~junction ~and ~on ~all ~wires}~~ j \ne i ~. \non \\
& &
\eea

We introduce an on-site potential which is equal to $\mu$ at all sites. In the
absence of interactions, the ground state is one in which all the states with 
energies going from $=-2t$ up to $\mu$ are filled; the Fermi wave number $k_F$
is given by $\mu = - 2 t \cos k_F$, assuming that $\mu$ lies in the range 
$[-2t, 2t]$. [We can then redefine all the wave numbers $k$ by subtracting 
$k_F$ from them as indicated after Eq. (\ref{wavefn}); the redefined wave 
numbers then run from $-\La$ to $\La$, where $\La$ is of order $k_F$.]

Let us now consider coupling the impurity spin to the sites labeled as
$n=1$ on the different wires by the following Hamiltonian
\bea
H_{\rm spin} &=& F_1 ~{\vec S} \cdot ~\sum_i ~\sum_{\al ,\be} ~\Psi_\al^\dg 
(i,1) ~\frac{{\vec \si}_{\al \be}}{2} ~\Psi_\be (i,1) \non \\
& & + ~F_2 ~{\vec S} \cdot ~\sum_{i\ne j} ~\sum_{\al ,\be} ~\Psi_\al^\dg 
(i,1) ~\frac{{\vec \si}_{\al \be}}{2} ~\Psi_\be (j,1) ~, \non \\
& &
\label{hspin3}
\eea
where $\Psi_\al (i,1)$ denotes the second quantized electron field at site 1 
on wire $i$ with spin $\al$. (Eq. (\ref{heff2}) below will provide a 
justification for this Hamiltonian). In Eq. (\ref{hspin3}), $F_1$ and $F_2$ 
denote amplitudes for spin-dependent scattering from the impurity within the 
same wire and between two different wires respectively.
Namely, a spin-up electron coming in through one wire can get scattered by the
impurity spin as a spin-down electron either along the same wire ($F_1$) or 
along a different wire ($F_2$). We then find that the Kondo coupling 
matrix $J_{ij}$ in Eq. (\ref{hspin2}) is as follows: all the diagonal entries 
are given by $J_1$ and all the off-diagonal entries are given by $J_2$, where 
\bea
J_1 &=& 4 F_1 \sin^2 k_F ~, \non \\
{\rm and} ~~~J_2 &=& 4 F_2 \sin^2 k_F
\label{jab1}
\eea
for modes with redefined wave numbers lying close to zero. 
This is precisely the kind of Kondo matrix whose RG flows 
were studied in Sec. IV. A. The flows of the parameters $J_1$ and $J_2$ 
considered there can be translated into flows of the parameters $F_1$ and 
$F_2$ here. In particular, the approach to the FP at $(J_1,J_2)=(0,0)$ given 
by Eq. (\ref{lowt1}) at low temperatures implies that spin-flip scattering 
within the same wire or between two different wires will have quite different
temperature dependences.

The flows to strong coupling shown in Fig. 3 can be interpreted as follows.
In the first quadrant of Fig. 3, we see that $J_1 + (N-1) J_2$ goes to 
$\infty$ faster than $|J_1 - J_2 |$; Eq. (\ref{jab1}) then implies that $F_1$
and $F_2$ go to $\infty$. In the fourth quadrant of Fig. 3, $J_1 - J_2$ goes 
to $\infty$ faster than $|J_1 + (N-1) J_2 |$; this implies that $F_1$ goes to
$\infty$ and $F_2$ goes to $-\infty$ as $-F_1/(N-1)$.

These flows to strong coupling have the following interpretations. In the 
first case, $F_1$ and $F_2$ flow to $\infty$. From Eq. (\ref{hspin3}), this 
implies that the impurity spin (of magnitude $\cal S$) is strongly and 
antiferromagnetically coupled to only one electronic field, namely, the `centre
of mass' field given by $\sum_i \Psi (i,1) /\sqrt{N}$ (suppressing the spin 
labels and the Pauli matrices for the moment). Hence that field and the 
impurity spin will combine to form an effective spin of ${\cal S} - 1/2$. In 
analogy with the three-dimensional Kondo problem, we can say that the impurity
spin is under-screened or exactly screened if ${\cal S} > 1/2$ or $ {\cal S} =
1/2$ respectively. In the second case, $F_1$ and $F_2 = -F_1/(N-1)$ go to 
$\infty$. Using Eq. (\ref{hspin3}), we can then show that the impurity spin 
is strongly and antiferromagnetically coupled to the $N-1$ `difference' fields
(given by the orthogonal combinations $[\Psi (1,1) - \Psi (2,1)]/\sqrt{2}$, 
$[\Psi (1,1) + \Psi (2,1) - 2 \Psi (3,1)]/\sqrt{6}, \cdots$). Hence those 
fields and the impurity spin will combine to give an effective spin of 
${\cal S} - (N-1)/2 = {\cal S} + 1/2 - N/2$. Thus the impurity spin is 
under-screened, exactly screened or over-screened if $2{\cal S}+1$ is 
greater than, equal to or less than $N$ respectively.

\subsection{\bf B. Griffiths $S$-matrix for $N$ wires}

This system can again be realized by the lattice shown in Fig. 6 (a) and
a tight-binding Hamiltonian. However, we now take the hopping amplitude to be
$-t$ on all bonds, except for the bonds which connect the sites labeled as
$n=1$ on each wire to the junction site; on those bonds, we take the hopping 
amplitude to be $t_1 = -t \sqrt{2/N}$. The on-site potential is taken to be 
$\mu$ at all sites, including the junction. We then find that the $S$-matrix
is of the Griffiths form for all values of the wave number $k$. Namely, for 
a wave which is incoming on wire $i$ with a wave number $k$, the wave function
is given by
\bea
\psi_{ik} (n) &=& e^{-ikn} ~-~ (1 - \frac{2}{N})~ e^{ikn} ~~{\rm on ~wire~} 
i ~, \non \\
&=& \frac{2}{N} ~e^{ikn} ~~{\rm on ~all ~wires}~~ j \ne i ~, \non \\
&=& \frac{2}{N} ~~{\rm at ~the ~junction ~site} ~.
\eea

We now consider coupling the impurity spin to the junction site labeled
by zero, and the sites labeled as $n=1$ on the different wires by the 
following Hamiltonian
\bea
H_{\rm spin} &=& F_3 ~{\vec S} \cdot ~\sum_{\al ,\be} ~\Psi_\al^\dg (0) ~
\frac{{\vec \si}_{\al \be}}{2} ~\Psi_\be (0) \non \\
& & + ~F_4 ~{\vec S} \cdot ~\sum_i ~\sum_{\al ,\be} ~\Psi_\al^\dg (i,1) ~
\frac{{\vec \si}_{\al \be}}{2} ~\Psi_\be (i,1) ~, \non \\
& &
\label{hspin4}
\eea
where $\Psi_\al (0)$ denotes the second quantized electron field at the
junction site with spin $\al$. (Sec. VI will provide a justification for 
this kind of a coupling). Then the Kondo coupling matrix $J_{ij}$ in Eq. 
(\ref{hspin2}) takes the following form: all the diagonal entries are given 
by $J_1$ and all the off-diagonal entries are given by $J_2$, where
\bea
J_1 &=& \frac{4F_3}{N^2} ~+~ 2F_4 ~[~ 1 ~-~ (1 - \frac{2}{N}) \cos 2k_F ~] ~,
\non \\
{\rm and} ~~~J_2 &=& \frac{4F_3}{N^2} ~+~ \frac{4F_4}{N} \cos 2k_F
\eea
for modes with wave numbers lying close to zero. The RG flows of this 
kind of Kondo matrix were studied in Sec. IV. B.

In terms of $F_3$ and $F_4$, the variables in Eq. (\ref{asymp}) are given by
\bea
J_1 ~-~ J_2 &=& 2 F_4 ~(1 - \cos 2k_F)~, \non \\
{\rm and} ~~~ J_1 + (N-1) J_2 &=& \frac{4F_3}{N} ~+~ 2F_4 ~(1 + \cos 2k_F) ~. 
\non \\
& &
\label{jab2}
\eea
Since $0 < k_F < \pi$, $1 \pm \cos 2k_F$ lie between 0 and 2. In the first 
quadrant of Fig. 4, $J_1 + (N-1) J_2$ goes to $\infty$ faster than
$|J_1 - J_2 |$; Eq. (\ref{jab2}) then implies that $F_3$ goes to $\infty$ and
$|F_4 | \ll F_3$. In the fourth quadrant of Fig. 4, $J_1 - J_2$ goes to 
$\infty$ faster than $|J_1 + (N-1) J_2 |$; this implies that $F_4$ goes to 
$\infty$ and $F_3$ goes to $-\infty$. 

These flows to strong coupling have the following interpretations. In the first
case, $F_3$ flows to $\infty$ which means that the impurity spin (of magnitude
$\cal S$) is strongly and antiferromagnetically coupled to an electron spin at
the junction site $n=0$; hence those two spins will combine to form an 
effective spin of ${\cal S} - 1/2$. (This case will be discussed in detail in 
Sec. VI). In the second case, $F_3$ goes to $-\infty$ while $F_4$ goes to 
$\infty$; hence the impurity spin is coupled strongly and ferromagnetically 
to an electron spin at the site $n=0$, and antiferromagnetically to electron
spins at the sites labeled as $n=1$ on each of the $N$ wires (see Fig. 6 (a) 
for the site labels). Hence the impurity spin will combine with those $N+1$
spins to form an effective spin of ${\cal S} + 1/2 - N/2$. Interestingly,
we see that the magnitudes of the effective spins formed in the strong 
coupling limits in the first and fourth quadrants are the same in the cases 
of $N$ disconnected wires and the Griffiths $S$-matrix. 

\subsection{\bf C. Chiral $S$-matrix for three wires}

This system can be realized by a lattice of the form shown in Fig. 6 (b).
The three wires meet at a triangle; the sites on each wire are labeled
as $n=1,2,\cdots$. The hopping amplitude is taken to be $-t$ on all the
bonds, except for the three bonds on the triangle. On those bonds, we take
the hopping amplitude to be complex, and of the form $-t e^{i\theta}$ in the
clockwise direction and $-t e^{-i\theta}$ in the anticlockwise direction. 
[We can think of the total phase $3\theta$ of the product of hopping 
amplitudes around the triangle as being the Aharonov-Bohm phase
arising from a magnetic flux enclosed by the
triangle. Such a flux breaks time reversal symmetry which makes the $S$-matrix
non-symmetric. Note that since only the value of $3\theta$ modulo $2\pi$ has 
any physical significance, we are free to shift the value of $\theta$ by 
$\pm 2\pi /3$. This changes the phase of the coupling $J_2$ defined below.] 
We then find that the $S$-matrix is of the chiral form given 
by Eq. (\ref{chirs}), provided that the wave number $k$ satisfies 
\beq
e^{i (3 \theta + k)} ~=~ - ~1 ~.
\label{thet}
\eeq
The phase $\gamma$ in (\ref{chirs}) is then given by $e^{i(\theta + k)}$. 
[Unlike the disconnected and Griffiths cases, we have not found a
lattice model which gives an $S$-matrix as in (\ref{chirs}) for {\it all} 
values of the wave number $k$.] Given a value of $\theta$, we therefore choose
a chemical potential $\mu = - 2 t \cos k_F$ such that $k_F$ satisfies Eq. 
(\ref{thet}). Since the properties of a fermionic system at low temperatures 
are governed by the modes near $k_F$, the above prescription produces a system
with a chiral $S$-matrix.

We now consider coupling the impurity spin to the three sites of the triangle
through the Hamiltonian
\bea
H_{\rm spin} &=& ~F_5 ~{\vec S} \cdot ~\sum_i ~\sum_{\al ,\be} ~\Psi_\al^\dg 
(i,1) ~ \frac{{\vec \si}_{\al \be}}{2} ~\Psi_\be (i,1) ~. \non \\
& &
\label{hspin5}
\eea
Then the Kondo coupling matrix $J_{ij}$ in Eq. (\ref{hspin2}) takes the 
form given in Eq. (\ref{konchi}), where
\bea
J_1 &=& 2F_5 ~, \non \\
{\rm and} ~~~J_2 &=& F_5 e^{-i(\theta + 3k_F)}
\eea
for modes with wave numbers lying close to zero. This is a special case of
the Kondo matrix given in Eq. (\ref{konchi}). [To obtain the most general
form given in (\ref{konchi}), we need to introduce another parameter, such 
as a coupling of the impurity spin to the sites labeled by $n=2$ in Fig.
6 (b).] The RG flows of this kind of Kondo matrix were studied in Sec. IV. C.

\section{\bf VI. Expansion around a strong coupling fixed point}

In Sec. V, we considered several examples of $S$-matrices and the RG flows
of the Kondo coupling. In most cases, we found that the Kondo couplings flow 
to large values. We can now ask whether the vicinity of the strong coupling 
FPs can be studied in some way. We will see that it is possible to do so 
through an expansion in the inverse of the Kondo coupling \cite{noz1}.

We will consider one example of such an expansion here. Following the 
discussion given after Eq. (\ref{jab2}), let us assume 
that the RG flows for the case of the Griffiths $S$-matrix have taken us to 
a strong coupling FP along the direction $J_2 /J_1 =1$, as shown
in the first quadrant of Fig. 4. This implies that the coupling 
of the impurity spin $S$ to an electron spin at the junction site $n=0$ has 
a large and positive (antiferromagnetic) value $F_3$, while its coupling to 
the sites labeled as $n=1$ on each of the wires has a finite value $F_4$ which
is much less than $F_3$ (the site labels are shown in Fig. 6 (a)). The ground 
state of the $F_3$ term (namely, just the first term in Eq. (\ref{hspin4}))
consists of a single electron at site $n=0$ which forms a total spin of 
${\cal S} - 1/2$ with the impurity spin. The energy of this spin state is 
$-F_3 ({\cal S}+1)/2$; this lies far below the high energy states in which 
there is a single electron at site $n=0$ which forms a total spin of 
${\cal S}+1/2$ with the impurity spin (these states have energy $F_3 
{\cal S}/2$), or the states in which the site $n=0$ is empty or doubly 
occupied (these states have zero energy). 

We can now do a perturbative expansion in $1/F_3$. We take the unperturbed
Hamiltonian to be one in which the hopping amplitudes on all the bonds are 
$-t$, except for the bonds connecting the sites labeled as $n=1$ on the 
different wires to the junction site; we take those hopping amplitudes to be 
zero. (This means that the unperturbed Hamiltonian corresponds to the case of 
$N$ disconnected wires). We also include the spin coupling proportional to 
$F_3$ in the unperturbed Hamiltonian. We take the perturbation $H_{\rm pert}$
as consisting of (i) the hopping amplitude $t_1$ on the bonds connecting the 
sites labeled as $n=1$ to the junction site, and (ii) the $F_4$ term in Eq. 
(\ref{hspin4}). Using this perturbation, we can
find an effective Hamiltonian \cite{noz1}. [Once again, we use the expression 
in Eq. (\ref{heff1}), where the high energy states are the ones listed in the 
previous paragraph. We will work up to second order in $t_1$ and $F_4$.] If 
${\cal S} > 1/2$, we find that the effective Hamiltonian has no terms of 
order $t_1$ or $t_1 F_4$, and it is given by
\bea 
H_{\rm eff} &=& F_{\rm 1, eff} ~{\vec S}_{\rm eff} \cdot ~\sum_i ~{\vec s}_i
\non \\
& & +~ F_{\rm 2, eff} ~{\vec S}_{\rm eff} \cdot \sum_{i \ne j} ~
\sum_{\al ,\be} ~ \Psi^\dg_\al (i,1) ~\frac{{\vec \si}_{\al \be}}{2} ~
\Psi_\be (j,1) \non \\ 
& & +~ C ~\sum_{i \ne j} ~({\vec S}_{\rm eff} \cdot {\vec s}_i)~ 
({\vec S}_{\rm eff} \cdot {\vec s}_j) ~+~ D ~\sum_{i < j} {\vec s}_i \cdot 
{\vec s}_j \non \\
& &
\label{heff2}
\eea
plus some constants, where 
\bea
{\vec s}_i &=& \sum_{\al ,\be} ~\Psi_\al^\dg (i,1) ~
\frac{{\vec \si}_{\al \be}}{2} ~\Psi_\be (i,1) ~, \non \\
F_{\rm 1, eff} &=& - ~\frac{8 ~t_1^2}{({\cal S}+1) ~(2{\cal S}+1) ~F_3} ~+~ 
\frac{2({\cal S}+1)F_4}{2{\cal S}+1} \non \\
& & - ~\frac{2 ~({\cal S}+1) ~F_4^2}{(2{\cal S}+1)^3 ~F_3} ~, \non \\
F_{\rm 2, eff} &=& - ~\frac{8 ~t_1^2}{({\cal S}+1) ~(2{\cal S}+1) ~F_3} ~, 
\non \\
C &=& \frac{2 ~F_4^2}{(2{\cal S}+1)^3 ~F_3} ~, \non \\
{\rm and} ~~~D &=& - ~\frac{F_4^2}{(2{\cal S}+1) ~F_3} ~.
\eea
In Eq. (\ref{heff2}), ${\vec S}_{\rm eff}$ denotes an object with spin 
${\cal S}-1/2$. We thus find a weak interaction between the spin ${\cal S} 
- 1/2$ and all the sites which are nearest neighbors of the site $n=0$ as 
shown in Fig. 6 (a).

If the impurity is a spin-1/2 object (i.e., ${\cal S}=1/2$), then the electron
at the site $n=0$ forms a singlet with the impurity. In that case, only the
last term in Eq. (\ref{heff2}) survives. However, there are other terms
in the effective Hamiltonian which are of higher order in $t_1 /F_3$ than in 
(\ref{heff2}); these have been calculated in Ref. \cite{noz2} for
the case ${\cal S}=1/2$. One of these terms describes spin-independent 
tunneling from one wire to another, of the form $\sum_{i \ne j} \sum_\al 
\Psi^\dg_\al (i,1) \Psi_\al (j,1)$. This is a contribution to the $S$-matrix
at the junction, and it can contribute to the conductance from one wire to
another as we will discuss in Sec. VIII.

Returning to the case ${\cal S} > 1/2$, we note that the last two terms in Eq.
(\ref{heff2}) are irrelevant as boundary operators if $g_2 (L=\infty) 
/(2\pi v_F)$ is
small; this is because ${\vec s}_i$ has the scaling dimension $1 - 
g_2 /(2\pi v_F)$ (as one can see from Eq. (\ref{disc})), and therefore the 
product ${\vec s}_i \otimes {\vec s}_j$ has the scaling dimension $2(1 - 
g_2 /(2\pi v_F))$ which is larger than $1$. The first two terms in Eq. 
(\ref{heff2}) have the same form as in Eqs. (\ref{hspin3}) and (\ref{jab1}), 
where the effective Kondo couplings
\bea
J_{\rm 1, eff} &=& 4 F_{\rm 1, eff} ~\sin^2 k_F ~, \non \\
{\rm and} ~~~ J_{\rm 2, eff} &=& 4 F_{\rm 2, eff} ~\sin^2 k_F
\eea
are equal, negative and small. We can now study the RG flow of this as in Sec.
IV. A. With these initial conditions, Eq. (\ref{disc}) and Fig. 3 show that the
Kondo couplings flow to the FP at $(J_{\rm 1, eff}, J_{\rm 2, eff}) = (0,0)$.

In this example, therefore, we obtain a picture of the RG flows at both short 
and large length scales. We start with the Griffiths $S$-matrix with certain 
values of the Kondo coupling matrix, and we eventually end at the stable FP of
the disconnected $S$-matrix for repulsive interactions, $g_2 (L=\infty) > 0$.

We will not discuss here what happens for the other possible RG flow for the 
Griffiths $S$-matrix, in which $J_1$ and $J_2$ become large along the direction
$J_2 /J_1 = -1/(N-1)$. As we noted in Sec. V B, $N+1$ spins get coupled
strongly to the impurity spin in that case; an expansion in the inverse
coupling is much more involved in that case. For the same reason, we will
not discuss expansions in the inverse coupling for the flows to strong 
coupling for the disconnected and chiral $S$-matrices.

\section{\bf VII. Decoupled wires with strong interactions}

In this section, we will briefly discuss what happens if there are $N$ 
decoupled wires and the interactions are strong. For the decoupled $S$-matrix,
one can `unfold' the electron field in each semi-infinite wire to obtain a 
chiral electron field in an infinite wire, and then bosonize that chiral field
\cite{gogolin,rao,giamarchi}. In the language of bosonization, the interaction
parameters are given by $K_\rho$ for the charge sector and $K_\si$ for the 
spin sector. Spin rotation invariance implies that $K_\si =1$, while $K_\rho$
is related to our parameters $g_i$ as follows \cite{giamarchi},
\bea
K_\rho &=& \sqrt{\frac {1 ~+~ g_4/\pi v_F ~+~ (g_1-2g_2)/2\pi v_F}
{1 ~+~ g_4/\pi v_F ~-~ (g_1-2g_2)/2\pi v_F}} \non \\
& \to & ~1 ~+~ \frac{g_1-2g_2}{2\pi v_F} ~.
\label{krho}
\eea
In the second line of the above equation, we have taken the limit of small
$g_i$ since we have worked to lowest order in the $g_i$ in the earlier 
sections. From Eq. (\ref{g124}), we see that $2g_2 - g_1$ is invariant
under the RG flow. The case of repulsive interactions corresponds to
$2g_2 - g_1 > 0$, i.e., $K_\rho < 1$.

The case of two decoupled wires ($N=2$) has been studied by Fabrizio and 
Gogolin in Ref. \cite{fabrizio}. They showed that if the interactions are weak
enough, the Kondo couplings $J_1$ and $J_2$ are both relevant; their results 
then agree with those discussed in Sec. IV A. But if the interactions are 
sufficiently strong, i.e., $K_\rho < 1/2$, then $J_2$ is irrelevant and flows 
to zero.

We will now show that their results can be generalized to the case of $N$ 
wires; one finds that there is again a value of $K_\rho$ below which $J_2$ 
is irrelevant. Following Ref. \cite{fabrizio}, we can write the spin-up and 
down Fermi fields $\Psi_{i,\al}$ in wire $i$ in terms of the charge and spin 
bosonic fields $\Phi_{i,\rho}$ and $\Phi_{i,\si}$. Close to the junction 
denoted as $x_j = 0$, we have
\bea
\Psi_{i,\up} &\sim & \frac{\eta_{i,\up}}{\sqrt{2\pi d}} ~e^{i (\Phi_{i,\rho}
/\sqrt{2 K_\rho} ~+~ \Phi_{i,\si} / \sqrt{2} )} ~, \non \\
{\rm and} ~~~\Psi_{i,\dn} &\sim & \frac{\eta_{i,\dn}}{\sqrt{2\pi d}} ~
e^{i (\Phi_{i,\rho} /\sqrt{2 K_\rho} ~-~ \Phi_{i,\si} / \sqrt{2} )} ~,
\label{boson}
\eea
where we have used the fact that $K_\si = 1$, and we have not explicitly 
written the arguments of the fields ($x_i = 0$) for notational convenience. 
The $\eta_{i,a}$ denote Klein factors, and $d$ is a short distance cut-off;
these will not play any role below. 

In bosonic language, the Hamiltonian $H = H_0 + H_{\rm int}$ in Eqs. 
(\ref{h0}) and (\ref{hint2}) is given by 
\beq
H ~=~ \frac{1}{4\pi} ~\sum_i ~\int_0^\infty ~dx_i ~[~ v_\rho \left( 
\frac{\partial \Phi_{i,\rho}}{\partial x_i} \right)^2 ~+~ v_\si \left( 
\frac{\partial \Phi_{i,\si}}{\partial x_i} \right)^2 ]~ ,
\label{h12}
\eeq
where $v_\rho$, $v_\si$ denote the charge and spin velocities respectively.
The bosonic fields satisfy the commutation relations
\beq
[~ \frac{\partial \Phi_{i,a} (x_i)}{\partial x_i} ~,~ \Phi_{j,b} (x_j) ~] ~=~
i ~2 \pi ~\delta_{ab} ~\delta_{ij} ~\delta (x_i ~-~ x_j) ~,
\eeq
where $a,b = \rho , \si$. 

The impurity spin part of the Hamiltonian is given by
\bea
H_{\rm spin} 
&=& ~J_1 ~{\vec S} \cdot ~\sum_i ~\sum_{\al ,\be} ~\Psi_{i,\al}^\dg ~
\frac{{\vec \si}_{\al \be}}{2} ~\Psi_{i,\be} \non \\
& & + ~J_2 ~{\vec S} \cdot ~\sum_{i\ne j} ~\sum_{\al ,\be} ~\Psi_{i,\al}^\dg ~
\frac{{\vec \si}_{\al \be}}{2} ~\Psi_{j,\be} ~.
\label{hspin6}
\eea
The spin densities on different wires are given by
\beq
\frac{1}{2} ~[~ \Psi_{i,\up}^\dg \Psi_{i,\up} ~-~ \Psi_{i,\dn}^\dg 
\Psi_{i,\dn} ~]~ =~ \frac{1}{2\sqrt{2} \pi} ~ \frac{\partial 
\Phi_{i,\si}}{\partial x_i} ~.
\label{psi1}
\eeq
The other terms take the form
\bea
\Psi_{i,\up}^\dg \Psi_{i,\dn} &\sim & e^{-i \sqrt{2} \Phi_{i,\si}} ~, \non \\
\Psi_{i,\up}^\dg \Psi_{j,\up} &\sim & e^{(i/ \sqrt{2}) [ -\Phi_{i,\rho} /
\sqrt{K_\rho} - \Phi_{i,\si} + \Phi_{j,\rho} / \sqrt{K_\rho} + 
\Phi_{j,\si} ] ~} ~, \non \\
\Psi_{i,\up}^\dg \Psi_{j,\dn} &\sim & e^{(i/ \sqrt{2}) [ -\Phi_{i,\rho} /
\sqrt{K_\rho} - \Phi_{i,\si} + \Phi_{j,\rho} / \sqrt{K_\rho} - 
\Phi_{j,\si} ] ~} ~, \non \\
& &
\label{psi2}
\eea
and so on. In (\ref{hspin6}) and (\ref{psi2}), we have not explictly written 
the arguments of the fields, $x_i = x_j = 0$; we will continue to do this 
wherever convenient. [The bosonic forms of the fermion bilinears in Eqs. 
(\ref{psi1}) and (\ref{psi2}) are so different because we are using abelian 
bosonization. For the same reason, we will find it useful to distinguish 
between the different components of $J_1$ and $J_2$, i.e., $J_{1z}$, 
$J_{1\perp}$, $J_{2z}$, and $J_{2\perp}$.] Let us define $N$ `orthonormal' 
linear combinations of the spin boson fields, namely, the `centre of mass'
combination
\beq
\Phi^0_\si ~=~ \frac{1}{\sqrt{N}} ~\sum_i ~\Phi_{i,\si} ~,
\eeq
and the `difference' fields
\beq
\Phi^n_\si ~=~ \frac{1}{\sqrt{n(n+1)}} ~[~ \sum_{m=1}^n ~\Phi_{m,\si} ~-~ 
n ~\Phi_{n+1,\si} ~] ~,
\eeq
where $n=1,2,\cdots, N-1$. We can now write Eq. (\ref{hspin6}) in the 
bosonic language. We obtain
\bea
H_{\rm spin} &=& \frac{J_{1z}}{2\sqrt{2} \pi} ~S^z ~\sum_i ~\frac{\partial
\Phi_{i,\si}}{\partial x_i} \non \\
& & + \frac{J_{1\perp}}{4\pi d} ~[~ S^+ ~e^{i\sqrt{2/N} \Phi^0_\si} ~
\sum_i ~e^{i\sum_n a^n_i \Phi^n_\si} ~+~ H. c. ] \non \\
& & - \frac{J_{2z}}{\pi d} ~S^z ~\sum_{i<j} ~\sin \left(\sum_n b^n_{ij} 
\Phi^n_\si \right) ~ \non \\
& & ~~~~~~~~~~~~~~\times ~\sin \left( \frac{\Phi_{i,\rho} - 
\Phi_{j,\rho}}{\sqrt{2K_\rho}} \right) \non \\
& & + \frac{J_{2\perp}}{2\pi d} ~[~ S^+ ~e^{i\sqrt{2/N} \Phi^0_\si}~ \non \\
& & ~\times \sum_{i<j} ~e^{i \sum_n c^n_{ij} \Phi^n_\si} \cos \left( 
\frac{\Phi_{i,\rho} - \Phi_{j,\rho}}{\sqrt{2K_\rho}} \right) + H. c. ],
\non \\
& &
\label{hspin7}
\eea
where the sums over $n$ in the second, third and last lines
run over the `difference' fields $\Phi^n_\si$. The constants $a^n_i$, 
$b^n_{ij}$ and $c^n_{ij}$ in Eq. (\ref{hspin7}) satisfy the relations
\bea
\sum_n ~(a^n_i)^2 &=& 2 ~-~ \frac{2}{N} ~, \non \\
\sum_n ~(b^n_{ij})^2 &=& 1 ~, \non \\
{\rm and} ~~~\sum_n ~(c^n_{ij})^2 &=& 1 ~-~ \frac{2}{N}
\label{sum}
\eea
for all values of $i, j$.

We can remove the phase factors $\exp (i\sqrt{2/N} \Phi^0_\si)$ in Eq. 
(\ref{hspin7}) by performing an unitary transformation of the total Hamiltonian 
$H_{\rm tot}$ given by the sum of (\ref{h12}) and (\ref{hspin6}), namely,
$H_{\rm tot} \to U H_{\rm tot} U^\dg$ \cite{emery}, where
\beq
U ~=~ e^{-i S^z \sqrt{2/N} \Phi^0_\si} ~.
\eeq
After this transformation, Eq. (\ref{hspin7}) takes the form
\bea
H_{\rm spin} &=& \frac{\lambda}{2\sqrt{2} \pi} ~S^z ~\sum_i ~\frac{\partial
\Phi_{i,\si}}{\partial x_i} \non \\
& & + \frac{J_{1\perp}}{4\pi d} ~[~ S^+ ~\sum_i ~e^{i\sum_n a^n_i \Phi^n_\si}~
+~ H. c. ~] \non \\
& & - \frac{J_{2z}}{\pi d} ~S^z ~\sum_{i<j} ~\sin \left( \sum_n b^n_{ij} 
\Phi^n_\si \right) ~ \non \\
& & ~~~~~~~~~~~~~~\times ~\sin \left( \frac{\Phi_{i,\rho} - 
\Phi_{j,\rho}}{\sqrt{2K_\rho}} \right) \non \\
& & + \frac{J_{2\perp}}{2\pi d} ~[~ S^+ \sum_{i<j} ~e^{i \sum_n c^n_{ij} 
\Phi^n_\si} \cos \left( \frac{\Phi_{i,\rho} - \Phi_{j,\rho}}{\sqrt{2K_\rho}} 
\right) \non \\ 
& & ~~~~~~~~~~+~ H. c. ~] ~,
\label{hspin8}
\eea
where $\lambda = J_{1z} - 4 \pi v_\si /N$. We can now study the problem in the
vicinity of the point $\lambda = J_{1\perp} = J_{2z} = J_{2\perp} = 0$.
Note that this is a strong coupling FP, since $\lambda = 0$ implies that
\beq
J_{1z} ~=~ \frac{4 \pi v_\si}{N} ~.
\eeq
Since the scaling dimension of $e^{i\beta \Phi_{i,a}}$ is given by $\be^2 /2$,
for $a=\rho, \si$, we see from Eq. (\ref{sum}) that the operators multiplying
$J_{1\perp}, J_{2z}$ and $J_{2\perp}$ in Eq. (\ref{hspin8}) have the scaling
dimensions $1 - 1/N$, $1/2 + 1/(2K_\rho)$ and $1/2 - 1/N + 1/(2K_\rho)$
respectively. This impies that the $J_{1\perp}$ operator is always relevant,
while the $J_{2z}$ operator is irrelevant if $K_\rho < 1$ (repulsive 
interactions). Most interestingly, the $J_{2\perp}$ operator is relevant or
irrelevant depending on whether $K_\rho >$ or $< N/(N+2)$. For $N=2$, this
gives the critical value of $K_\rho$ to be 1/2 \cite{fabrizio}, while for
$N \to \infty$, the critical value of $K_\rho$ approaches 1, i.e., the limit
of weak repulsive interactions. 

We saw in Sec. IV A that a flow to strong coupling is indeed possible along 
the line $J_2 =0$, although that line is unstable to small perturbations in 
$J_2$. We now see that the line is stabilized (to first order in the
couplings) if the interactions are sufficiently strong, i.e., if
\beq
K_\rho ~<~  \frac{N}{N+2} ~.
\eeq
If $J_2$ flows to zero and $J_1$ flows to large values, Eq. (\ref{hspin3}) 
shows that the impurity spin is coupled strongly and antiferromagnetically to
the electron fields $\Psi (i,1)$ on all the $N$ wires; hence they will combine
to form an effective spin of ${\cal S} - N/2$. (If ${\cal S} < N/2$, the
impurity spin is over-screened). This describes a $N$-channel AFM FP 
with no coupling between channels \cite{oreg,pustilnik2}.

\section {\bf VIII. Conductance calculations}

Our calculations for the Kondo couplings can be explicitly applied to 
various geometries of quantum wires and a quantum dot (containing the
impurity spin) shown in Fig. 2, such as (a) a dot coupled independently to 
each wire (disconnected $S$-matrix for the wires), so that the conductance 
can only occur through the dot, or (b) a side-coupled dot (Griffiths 
$S$-matrix for the wires), where the conductance can occur directly between 
the wires. In general, of course, one can have any $S$-matrix at the junction,
so that the conductance can occur both through the dot and directly between 
the wires.

Let us now consider the conductance near the different FPs 
\cite{durga2,pustilnik1} for the case of weak interactions. In the Griffiths 
case where the conductance can occur directly between the wires, let us 
consider the case of small values of $J_1, J_2$ (both much smaller than
$2\pi v_F$), and $g_2 \gg g_1$. At high temperatures, before the $J_i$'s have
grown very much under RG, we see from Eq. (\ref{grif}) that $J_1$ remains 
small, while $J_2$ grows due to the term $g_2 J_2$. Namely, $J_2 \sim 
(T/T_K )^{-\nu}$, where $\nu = g_2 (L=\infty)/(2 \pi v_F)$. The effect of 
$J_2$ is to scatter the electrons from the impurity spin, and thereby 
reduce the conductance between any two wires from the maximal value of 
$G_0 = (4/N^2) e^2 /h$. Since the scattering probability is proportional to 
$J_2^2$, the conductance at high temperatures ($T \gg T_K$) is given by 
\beq
G ~-~ G_0 ~\sim ~- ~G_0 ~{\cal S} ({\cal S} +1) ~(T/T_K)^{-2\nu} ~.
\label{cond1}
\eeq
[The factor of ${\cal S} ({\cal S} +1)$ appears for the following reason.
Consider an electron coming in through wire $i$; it can have
spin up or down, and the impurity spin can have any value of $S^z$ from
${\cal S}$ to $-{\cal S}$. We assign all these $2(2{\cal S}+1)$ states the
same probability. As a result of the Kondo coupling $J_2$, the electron can
scatter to a different wire $j$; as a result, its spin may or may not flip, 
and the value of $S^z$ for the impurity spin can also change by 0 or $\pm1$. 
If we calculate the probabilities of all the different possible processes 
and add them, we get a factor of ${\cal S} ({\cal S} +1)$]. 
Using Eq. (\ref{krho}), we see that (\ref{cond1}) takes the form
\beq
G ~-~ G_0 ~\sim ~- ~G_0 ~{\cal S} ({\cal S} +1) ~(T/T_K)^{K_\rho-1} ~,
\label{cond2}
\eeq
where we have used the RG flow to set $g_1=0$ and $\nu = g_2 (L=\infty)/(2 
\pi v_F)$. On the other hand, if the leads were Fermi liquids ($g_1 = g_2 = 
0$), $J_2$ would be given by Eq. (\ref{lowt2}), and we would get 
\beq
G ~-~ G_0 ~\sim ~- ~\frac{G_0{\cal S}({\cal S} +1)}{\ln (T/T_K)^4} ~.
\label{cond3}
\eeq

At low temperatures, the Kondo couplings flow to large values; as discussed at
the end of Sec. VI, their behaviors are then governed by the FP at $(J_{1,{\rm
eff}}, J_{2,{\rm eff}}) = (0,0)$ of the disconnected wire case with an 
effective spin ${\cal S}_{\rm eff} = {\cal S} -1/2$. In this case, only 
$J_2^2$ contributes to the conductance between two different wires. From Eq. 
(\ref{lowt1}), we see that the conductance is given by
\bea
G &\sim & G_0 ~{\cal S}_{\rm eff} ({\cal S}_{\rm eff} +1) ~(T/T_K)^{2\nu} 
\non \\
&\sim & G_0 ~{\cal S}_{\rm eff} ({\cal S}_{\rm eff} +1) ~(T/T_K)^{1-K_\rho}
\label{cond4}
\eea
for $T \ll T_K$. For Fermi liquid leads, Eq. (\ref{lowt2}) implies that the 
conductance is given by
\beq
G ~\sim ~\frac{G_0{\cal S}_{\rm eff}({\cal S}_{\rm eff} +1)}{\ln (T/T_K)^4} ~.
\label{cond5}
\eeq
Thus a measurement of the temperature dependence of the conductance should be 
able to distinguish between the Fermi liquid and TLL cases at both high and 
low temperatures. For the case $N=2$, the expressions in Eqs. (\ref{cond2}) 
and (\ref{cond4}) agree with those given in Refs. \cite{furusaki,durga2}, but 
Eqs. (\ref{cond3}) and (\ref{cond5}) differ from the expressions given in 
earlier papers (like Ref. \cite{durga2}) for the powers of $1/\ln (T/T_K)$. 
(As we had discussed earlier after Eq. (\ref{lowt2}), we would get the same 
powers of $1/\ln (T/T_K)$ as in Ref. \cite{durga2} if $J_2$ was exactly equal 
to $J_1$ or $-J_1/(N-1)$).

The above expressions for the conductance shows that for both Fermi liquid 
leads and TLL leads (with repulsive interactions), and for both $T \gg T_K$ 
and $T \ll T_K$, the conductance increases with the temperature. It is then
natural to assume that this would be true for intermediate temperatures as 
well, so that the conductance increases monotonically with temperature from 
0 to $G_0$; this would be consistent with the results in Refs. 
\cite{furusaki,durga2}. It may be useful to discuss here why there is no 
Kondo resonance peak in the conductance at low temperatures in our model, in 
contrast to what is found in other models (for instance Refs. 
\cite{pustilnik1,glazman,ng}) and observed experimentally 
\cite{goldhaber,wiel}.
In our model, once the impurity spin gets very strongly coupled to the 
junction site in Fig. 6 (b) (due to the flow to large $J_1$ and $J_2$ in the 
Griffiths case), that site decouples from the wires; this leaves no other 
pathway for the electrons to transmit from one wire to another. In contrast to
this, if the junction region was more complicated (for instance, if there were
additional bonds which connect different wires without going through the 
impurity spin, or there was a dot with several energy levels through which 
the electron can transmit), then the electron may still be able to transmit 
even after the impurity quenches the electron on a single site (or level). 
Hence, it may be possible for the conductance to increase to the unitarity 
limit at the lowest temperatures; this is known to occur for models with Fermi
liquid leads. For TLL leads, however, our analysis remains valid even if there
are additional bonds between the wires, because any such direct tunneling 
amplitudes are irrelevant and renormalize to zero as shown in Eq. (\ref{tau}).

Finally, let us briefly consider the case of strong inter-electron 
interactions. For $K_\rho < N/(N+2)$, we saw in Sec. VII that a multi-channel 
FP gets stabilized in the case of $N$ disconnected wires. To obtain the 
conductance at this point, we need to study the operators perturbing this 
point, similar to the analysis in Refs. \cite{pustilnik2,kim,durga2}; this 
has not yet been done. 

\section{\bf IX. Conclusions}

To summarize our results, we have studied systems of TLL wires which meet at a 
junction. The junction is described by a spin-independent $S$-matrix, and 
there is an impurity spin which is coupled isotropically to the electrons in 
the neighborhood of the junction. The $S$-matrix and the Kondo coupling matrix
$J_{ij}$ satisfy certain RG equations. We have studied the RG flows of the 
Kondo couplings for a variety of FPs of the $S$-matrices. Although the Kondo 
couplings generally grow large, one can sometimes study the system through
an expansion in the inverse of the coupling. This leads to a new system in 
which the effective Kondo couplings are weak; the RG flows of these effective 
couplings can then be studied. 

For example, at the fully connected or Griffiths $S$-matrix, we find that 
for a range of initial conditions, the Kondo couplings can flow to a strong 
coupling FP along the direction $J_2 /J_1 =1$, where their fate is decided by 
a $1/J$ analysis. This analysis then shows that the couplings flow to the 
FM FP of the disconnected $S$-matrix lying at $(J_{1, {\rm eff}}, 
J_{2, {\rm eff}})= (0,0)$. For this system, therefore, one obtains a 
description of the system at both short and large length scales.

For the case of disconnected wires
and repulsive interactions, there is a range of Kondo couplings which flow
towards a multi-channel FM FP at $(J_1,J_2) = (0,0)$. At low temperatures,
we find spin-flip scattering processes with temperature dependences which are 
dictated by both the Kondo effect and the inter-electron interactions.
It may be possible to observe such scatterings by placing a quantum dot with
a spin at a junction of several wires with interacting electrons.

For other initial conditions for the disconnected case, the Kondo couplings
flow towards the strong coupling FPs at $J_1, |J_2| \to \infty$. In
general, this is just the single channel strong coupling AFM FP. But
there is a special line where $J_1 \to \infty$ and $J_2 = 0$; this is the 
multi-channel AFM FP. The RG equations show that both $J_1$ and $J_2$ are
relevant around the weak coupling FP if the interactions are weak. However,
if the interactions are sufficiently strong (i.e., $K_\rho < N /(N+2)$), we 
find that $J_2 \to 0$, and the multi-channel FP gets stabilized.

Experiments are underway to look for multi-channel FPs, and proposals have 
been made for minimizing the couplings between channels using gate voltages 
\cite{oreg}. We suggest here that enhancing inter-electron interactions in 
the wires offers another way of reducing the inter-channel coupling and thereby
observing the effects of the multi-channel FP.

Finally, we have discussed the temperature dependences of the conductances
close to the disconnected and Griffiths $S$-matrices, and showed that this 
also provides a way to distinguish between Fermi liquid leads and TLL leads.

\vskip .5 true cm
\centerline{\bf Acknowledgments}
\vskip .5 true cm

S.R. thanks P. Durganandini for discussions.
V.R.C thanks CSIR, India for financial support. V.R.C. and D.S. thank DST, 
India for financial support under the project No. SP/S2/M-11/2000.

\end{document}